\documentclass[11pt]{article}

\usepackage[preprint]{acl}

\usepackage{times}
\usepackage{latexsym}

\usepackage[T1]{fontenc}

\usepackage[utf8]{inputenc}

\usepackage{microtype}

\usepackage{inconsolata}

\usepackage{graphicx}

\usepackage{algorithm}
\usepackage{algpseudocodex}
\usepackage{multirow}
\usepackage{amsmath}
\usepackage{graphicx}
\usepackage{subcaption}
\usepackage{comment}
\usepackage{makecell}
\usepackage{array}
\usepackage{tcolorbox}
\usepackage{enumitem}
\usepackage{amsfonts}       
\usepackage{booktabs}       
\usepackage{url}            
\usepackage{xcolor}         
\usepackage{hyperref}
\usepackage{tabularx}

\title{How Jailbreak Attacks Inform Safety Alignment:\\A Defender-Centric, Shapley-Based Evaluation of Jailbreak Contributions}

\author{
 \textbf{Yukai Zhou\textsuperscript{1}},
 \textbf{Feiyang Lu\textsuperscript{1}},
 \textbf{Xiaokai Mao\textsuperscript{2}},
 \textbf{Jinfei Liu\textsuperscript{2}},
 \textbf{Wenjie Wang\textsuperscript{1}$^{\dagger}$}
\\
 \textsuperscript{1}ShanghaiTech University, \textsuperscript{2}Zhejiang University,\\
  \{zhouyk12023, lufy2023, wangwj1\}@shanghaitech.edu.cn,\\
  \{xiaokaimao, jinfeiliu\}@zju.edu.cn
}

\begin{document}
\maketitle

\begin{abstract}
Jailbreak attacks on large language models are usually evaluated by attacker-centric metrics such as attack success rate (ASR), yet an attack that breaks a model is not necessarily useful for improving its safety.
We propose a defender-centric view of jailbreak evaluation, 
where attacks are evaluated by the downstream safety improvements they enable when used as red-teaming data for safety training.
Building on this view, we introduce \textbf{A-MESS} (\textbf{Minimal Effective Attack-Subset Selection}), a setting-agnostic framework for attributing and selecting jailbreak attacks from black-box subset utility observations. 
A-MESS estimates \textbf{AttackSHAP}, 
a Shapley-based score that attributes marginal utility to individual attacks
and selects compact attack subsets under user-specified budgets via greedy or surrogate-based optimization.
Across controlled utility landscapes and real LLM safety settings, we find that ASR rankings are weakly aligned with defender-centric utility, that AttackSHAP can be estimated accurately with limited utility queries, 
and that directly optimizing subsets yields stronger safety utility than attacker-centric or attribution-only selection.
These results suggest evaluating jailbreak attacks as resources for improving safety, not only as tools for breaking models.
\footnote{
A-MESS project page: 
\href{https://attackshap.github.io}{attackshap.github.io}
}
\end{abstract}


\section{Introduction}

Jailbreak attacks have become a central tool for stress-testing the safety of large language models (LLMs)~\cite{carlini2023aligned,wei2023jailbroken,jain2023baseline,yi2024jailbreak,chu2025jailbreakradar}.
By constructing adversarial prompts~\cite{zou2023universal,zhu2024autodan,liao2024amplegcg,zhou2025don,beyer2026samplingaware} or interaction strategies~\cite{chao2025jailbreaking,ren2025llms,andriushchenko2025does,huang2026obscure}, these attacks expose failure modes where a model's safety alignment can be bypassed. 
Existing jailbreak research therefore commonly evaluates attacks using attacker-centric metrics such as attack success rate (ASR), computational cost, universality, and transferability.
These metrics measure how effectively and practically an attack can compromise a model from the perspective of an attacker.

For LLM safety researchers, however, this perspective is incomplete. 
A jailbreak attack is useful for safety research not merely because it breaks an undefended model, but because it can reveal weaknesses that will help improve a defended one~\cite{yu2024robust,mo2024fight,fu2025short}. 
By shifting from \emph{attack success} to \emph{defense utility},
this motivates a defender-centric view of jailbreak evaluation:
attacks should be valued by their downstream safety utility contribution when used as red-teaming resources in defense pipelines.
Under this view, attack success is not the final objective, but an input to a broader safety-improvement process.

This shift exposes a practical decision problem. 
\textbf{\emph{
Which jailbreak attack(s) matter most when ``matter'' means improving downstream safety alignment, rather than maximizing attack success?
}}
This question cannot be answered by standalone attack-success metrics alone.
If ASR were a reliable proxy for defense utility contribution, ranking attacks by no-defense success would already identify the most useful attacks for improving safety.
However, the value of an attack may depend on the specific safety setting context,
including the defense pipeline, evaluation benchmark,  utility definition, and even the other attacks selected with it for defense at the same time.
Thus, the setting is highly heterogeneous, and even potentially proprietary.
We therefore model each setting by a black-box utility function \(v_\theta(S)\), which measures the safety utility of an attack subset \(S\) under a user-specified setting \(\theta\). 
Our goal is not to find a universally best jailbreak attack, but to provide a setting-agnostic framework that can be instantiated wherever such subset utilities can be observed.

\begin{figure*}[t]
    \centering
    \includegraphics[width=0.98\linewidth]{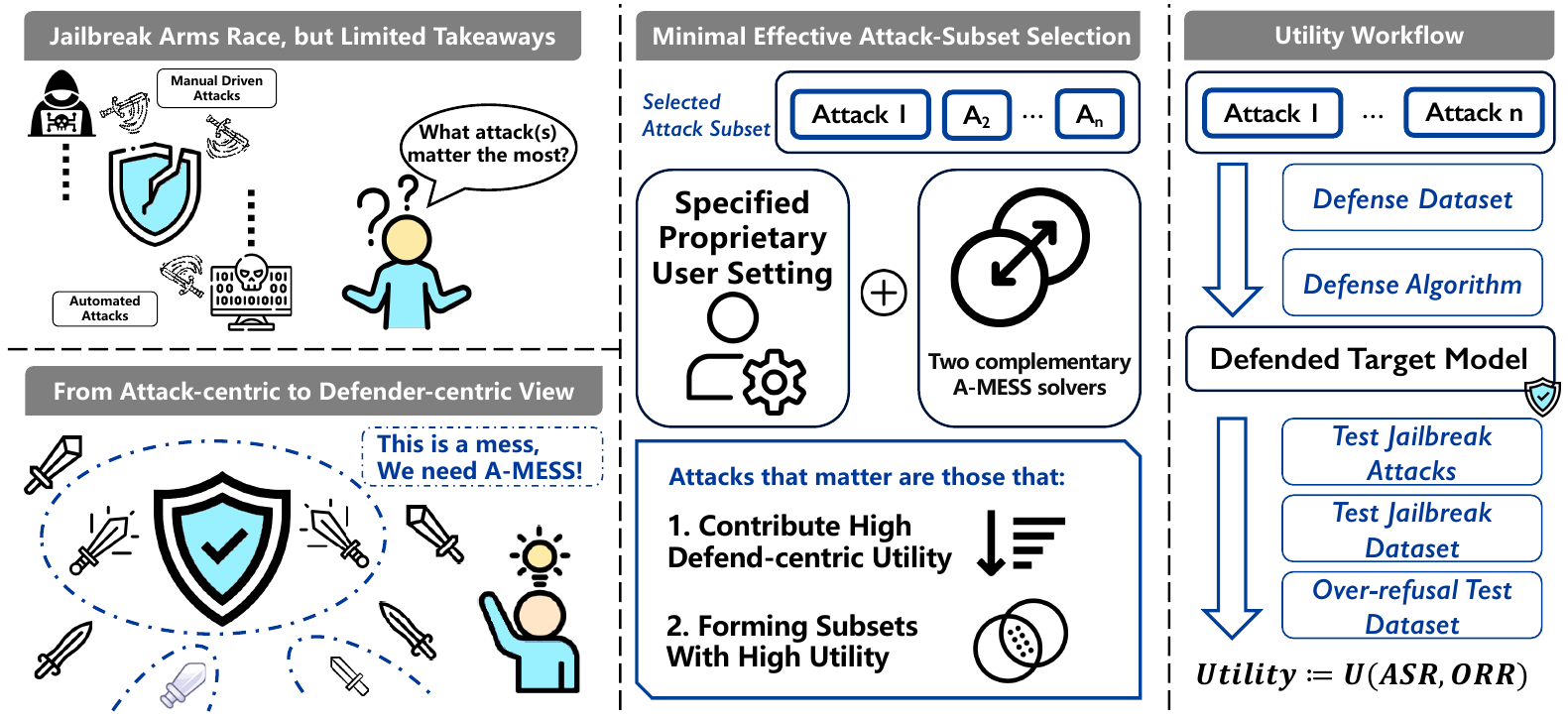}
    \caption{
    Conventional jailbreak research often follows an attacker-centric arms race, where attacks are primarily judged by how effectively they break a target model.
    A-MESS instead starts from a defender-centric question: which jailbreak attack(s) matter most for improving safety alignment under a specific, potentially proprietary user setting?
    Given setting-specific utility observations, A-MESS supports two complementary tasks:
    identifying attacks with high equitable contribution via AttackSHAP, and selecting compact attack subsets with high downstream utility.
    }
    \label{fig:placeholder}
    \vspace{-1em}
\end{figure*}

This decision problem formulation leads to two related but distinct tasks, as illustrated in Figure~\ref{fig:placeholder}.
The first is attribution: how much credit should each attack receive for contributing to downstream utility across different contexts? 
We address this with \textbf{AttackSHAP}, a Shapley-based attribution score that treats jailbreak attacks as players, attack subsets $S$ as coalitions, and \(v_\theta(S)\) as the coalition value. 
However, directly employing exact Shapley value computation requires the true subset utility landscape, 
which is prohibitively expensive to obtain.
Moreover, the same utility view also induces a subset-selection problem:
Under a limited budget, which subset should a defender actually use? 
These tasks are not equivalent, since attacks with high individual contribution do not necessarily compose the highest-utility subset.

To address these tasks,
we propose \textbf{A-MESS} (\textbf{Minimal Effective Attack-Subset Selection}), 
a defender-centric framework for both attack attribution and budgeted attack-subset selection tasks. 
A-MESS provides two complementary solvers. 
\textsc{A-MESS-SURROGATE} leverages a surrogate model to learn an approximate utility landscape from sampled subset evaluations, enabling efficient AttackSHAP estimation and multi-budget subset search under a fixed evaluation budget. 
\textsc{A-MESS-GREEDY} directly queries true utilities and greedily selects attacks with the largest marginal gains, providing a query-efficient option when a practitioner needs a single subset under a fixed budget.

\noindent Our contributions are threefold:
\begin{itemize}[leftmargin=15pt,itemsep=2pt,parsep=0pt,partopsep=0pt,topsep=0pt]

    \item We reframe jailbreak attack evaluation from attacker-centric success to defender-centric utility, and empirically test whether ASR predicts downstream safety utility.
    \item We introduce \textbf{AttackSHAP}, a Shapley-based attribution score for assigning setting-specific utility contribution to individual jailbreak attacks across subset contexts.
    \item We formulate budgeted attack-subset selection, 
    and propose \textbf{A-MESS} (\textbf{Minimal Effective Attack-Subset Selection}), 
    a setting-agnostic framework with greedy and surrogate-based solvers for selecting compact high-utility attack subsets from black-box utility observations.
    Empirical results show that A-MESS yields subset with higher defender-centric utility than attacker-centric or attribution-only selection.

\end{itemize}

\section{Preliminaries and Problem Setup}
\label{sec:preliminaries}

We first formalize the problem discussed above.
Let \(\mathcal{A}=\{a_1,\ldots,a_n\}\) denote the candidate attack universe, where each \(a_i\) is an attack algorithm, or equivalently attack-generated data source.
As summarized in Figure~\ref{fig:placeholder}, a safety setting \(\theta\) specifies how one attack subset is converted into a defended system and how that system is evaluated.
It may include the target model, defense procedure, evaluation benchmarks, computation budget, and utility definition.
For any subset \(S\subseteq\mathcal{A}\), we define a defender-centric utility
\(v_\theta:2^{\mathcal{A}}\to\mathbb{R}\), where \(v_\theta(S)\) is the utility obtained after applying the defense workflow with attacks in \(S\).
The utility may measure jailbreak ASR reduction, or combine it with other objectives such as over-refusal penalty.
A-MESS does not assume access to the internal details of \(\theta\), 
it only assumes that a practitioner can query \(v_\theta(S)\) for selected subsets.
Throughout the paper, we instantiate \(v_\theta\) in both real LLM safety settings and controlled synthetic utility landscapes. 
In each case, utilities are computed according to the corresponding setting, and normalized to a common \([0,1]\) scale. Detailed utility settings are provided in the experimental sections and Appendix.

In this formulation, a utility query \(v_\theta(S)\) corresponds to executing the defender's evaluation workflow for one attack subset: constructing defense data or prompts from \(S\), applying the defense procedure, and evaluating the resulting system under the setting-specific safety metrics.
This black-box view is important because many realistic safety pipelines stay proprietary, as model providers tend not to release the entire pipeline.
Attacker-centric quantities such as no-defense ASR can be used as baselines, but they do not explicitly query \(v_\theta(S)\).


\paragraph{Individual attack attribution.}
A central goal is to quantify how much each attack contributes to the safety utility. 
This cannot be captured by standalone attack metrics since they ignore subset context:
Two attacks may be redundant if they expose the same failure mode, so adding one after the other yields little additional utility.
Conversely, two individually modest attacks may be complementary if they cover different weaknesses and jointly improve the defended system.
AttackSHAP is designed to average such marginal contributions across subset contexts.
We therefore view attacks as players in a cooperative game,
attack subsets \begin{small}\(S\subseteq\mathcal{A}\)\end{small} 
as coalitions and \(v_\theta(S)\) as the coalition value.
The AttackSHAP value of attack \(a_i\) is then:

\begin{small}
\begin{equation}
\phi_i^\theta =
\sum_{S\subseteq\mathcal{A}\setminus\{a_i\}}
\frac{|S|!(n-|S|-1)!}{n!}
\left[v_\theta(S\cup\{a_i\})-v_\theta(S)\right].
\end{equation}
\end{small}

\noindent Or equivalently, let \begin{small}\(\Pi(\mathcal{A})\)\end{small} be the set of all permutations of attacks and \(P_i^\pi\) be the set of attacks appearing before \(a_i\) in permutation \(\pi\). The same value can be written in permutation form as

\begin{small}
\begin{equation}
\phi_i^\theta =
\frac{1}{n!}\sum_{\pi\in\Pi(\mathcal{A})}
\left[v_\theta(P_i^\pi\cup\{a_i\})-v_\theta(P_i^\pi)\right],
\end{equation}
\end{small}

\noindent This is the Shapley value~\cite{shapley1953value,dubey1975uniqueness,ghorbani2019data} of \(a_i\) in the utility game induced by \(\theta\),
and quantifies the average marginal utility of adding \(a_i\) across all possible
attack orderings.
We compare AttackSHAP with two boundary marginal baselines:
\(\mathrm{AOO}_i^\theta=v_\theta(\{a_i\})-v_\theta(\emptyset)\), which measures empty-context gain, and
\(\mathrm{LOO}_i^\theta=v_\theta(\mathcal{A})-v_\theta(\mathcal{A}\setminus\{a_i\})\), which measures full-context loss~\citep{koh2017understanding,ghorbani2019data}.
We compare estimated AttackSHAP against reference scores using standard regression metrics along with Pearson and Spearman correlations.

\paragraph{Budgeted subset selection.}
Attribution ranks individual attacks, but LLM safety researchers often need a subset-level decision in practice. Given a budget \(k\), the oracle objective is to select the size-\(k\) subset with the highest safety utility
\begin{equation}
S_k^\star = \arg\max_{S\subseteq\mathcal{A},\,|S|=k} v_\theta(S).
\end{equation}
Here \(S_k^\star\) denotes the oracle optimal subset under setting \(\theta\). 
This fixed-size formulation captures the common case where a defender can only use \(k\) attacks for training or evaluation.
In practice, finding \(S_k^\star\) exactly is infeasible for large \(n\), because it requires evaluating \(\binom{n}{k}\) candidate subsets. 
We use \(S_k^\star\) as an oracle upper bound when the full utility landscape is available, and introduce practical A-MESS solvers for approximating high-utility subsets \(S_k(\textsc{A-MESS})\) in Section~\ref{sec:method}.

\section{Method}
\label{sec:method}

We now describe the computational components of A-MESS. 
When the complete real utility landscape is too expensive to compute, we first learn a surrogate utility model \(\hat v_\theta\) from sampled subset evaluations, enabling efficient AttackSHAP estimation.
The second selects compact attack subsets under a user-specified budget, 
either by direct greedy search over observed true utilities or by search over the learned surrogate utility landscape.

\subsection{Surrogate Utility Learning}

\textsc{A-MESS-Surrogate} learns an approximate utility landscape from sampled subset evaluations. 
For each sampled training subset \(T_j\subseteq\mathcal{A}\), we represent it by a binary indicator vector \(x_{T_j}\in\{0,1\}^n\), where \((x_{T_j})_i=1\) iff \(a_i\in T_j\). Given \(M\) queried utilities, \(\mathcal{D}_{\mathrm{train}}=\{(x_{T_j},v_\theta(T_j))\}_{j=1}^M\), we train a surrogate model \(\hat v_\theta\) to predict subset utility from the indicator vector by minimizing mean squared error. Once trained, \(\hat v_\theta\) can be evaluated on unqueried subsets at low cost, enabling AttackSHAP estimation, multi-budget subset search, and repeated selection decisions 
without repeatedly querying the true utility function. 

In practice, we split the queried subset-utility observations into training, validation, and held-out test sets. 
The training split is used to fit \(\hat v_\theta\), the validation split is used for model hyper-parameter selection and early stopping, and the held-out test split evaluates whether the learned surrogate generalizes to unseen attack subsets. 
In synthetic landscapes, held-out subsets are sampled from the fully enumerated utility table, while in real LLM settings they correspond to additional observed subset utilities that are not used for training. 

\subsection{Attack Attribution and Ranking}

A-MESS uses AttackSHAP to estimate the contribution of each attack. When the true utility landscape is available, we compute reference scores \(\phi_i^\theta=\phi_i(v_\theta)\), either exactly in controlled synthetic settings or approximately through sampled coalitions. When using a surrogate model, \textsc{A-MESS-Surrogate} estimates \(\hat\phi_i^\theta=\phi_i(\hat v_\theta)\), replacing true utility evaluations with surrogate predictions.
Thus, after the initial utilities are queried to fit the surrogate model, additional coalitions required by Shapley estimation are evaluated by \(\hat v_\theta\) rather than by re-running the full defense workflow.


Attribution scores can also be used for top-\(k\) selection by ranking attacks and returning the highest-scoring \(k\) elements. We consider rankings induced by ASR, AttackSHAP, LOO~\cite{koh2017understanding}, and AOO. This allows us to separate two questions: whether utility can be accurately attributed to each individual attack, and whether selecting attacks by individual importance is sufficient for subset-level utility maximization.


\subsection{Budgeted Subset Selection}

Beyond ranking individual attacks, A-MESS directly addresses the budgeted subset selection problem. Given a subset size \(k\), the goal is to return a high-utility subset \(S_k(m)\) under the current setting. We consider two complementary solvers.

\paragraph{\textsc{A-MESS-Greedy}.}

The greedy solver directly queries \(v_\theta\) and greedily selects attacks by marginal gain, making it suitable for single-budget selection when true utility queries are affordable. 
It starts from \(S_0=\emptyset\) and iteratively adds the attack with the largest true marginal utility gain:

\begin{small}
\[
a_t^\star
=
\arg\max_{a\in\mathcal{A}\setminus S_t}
\bigl[v_\theta(S_t\cup\{a\})-v_\theta(S_t)\bigr],
\]
\vspace{-1em}
\[
S_{t+1}=S_t\cup\{a_t^\star\}.
\]
\end{small}

\noindent After \(k\) steps, it returns \(S_k(\textsc{A-MESS-Greedy})=S_k\). This solver is appropriate when the practitioner only needs one subset under a fixed budget and can afford adaptive true utility queries. 
Its query cost is \begin{small}
    \(\sum_{t=0}^{k-1}(n-t)=nk-k(k-1)/2\)
\end{small} for a size-\(k\) subset, excluding cached evaluations.

\paragraph{\textsc{A-MESS-Surrogate}.}
The surrogate solver first learns \(\hat v_\theta\) from sampled subset evaluations and then searches in the learned utility landscape. For a target budget \(k\), it returns

\begin{small}
\[
S_k(\textsc{A-MESS-Surrogate})
=
\arg\max_{S\subseteq\mathcal{A},\,|S|=k}
\hat v_\theta(S).
\]
\end{small}


\subsection{Defense-pipeline Instantiations}
A-MESS only requires black-box access to \(v_\theta(S)\), but in real LLM experiments we instantiate each utility query with two defense pipelines: adversarial training~\cite{mazeika2024harmbench} and in-context defense~\cite{wei2023jailbreak} (ICD). 
We distinguish defense jailbreak data from test data: 
the former is collected for the selected attack subset \(S\) and used to build the defended system, while the latter is held out for evaluating \(v_\theta(S)\). 
The resulting defended system is evaluated on held-out jailbreak tests, and on over-refusal tests when included by the utility definition.

\section{Experiments}

\subsection{Evaluated Aspects and Metrics}
We evaluate A-MESS along three aspects. 
First, we test whether attacker-centric success is aligned with defender-centric utility by comparing the no-defense ASR ranking with the singleton utility ranking.
We report Spearman's \(\rho\), Kendall's \(\tau\), and representative rank shifts. 
Second, we evaluate whether AttackSHAP can be estimated under limited utility queries. 
For synthetic landscapes, we compare estimated AttackSHAP against exact references using regression and rank-correlation metrics.
For real LLM utilities, where the full landscape is not exhaustively observed, we evaluate held-out prediction of \(v_\theta(S)\). 
Third, we evaluate budgeted subset selection using raw subset utility \(v_\theta(S_k(m))\) and Normalized Gain:

\begin{equation}
\mathrm{NG}_k(m)=
\frac{
v_\theta(S_k(m))-v_\theta(S_k(\textsc{Base}))
}{
v_\theta(S_k^\star)-v_\theta(S_k(\textsc{Base}))
}.
\end{equation}

\noindent Here \(S_k(\mathrm{BASE})\) is the average random size-\(k\) subset and \(S_k^\star\) is the oracle size-\(k\) subset when available. 
Thus, \(NG_k(m)\) measures the fraction of the oracle improvement gap recovered by method \(m\).

\subsection{Experimental Setup}


\paragraph{Synthetic utility landscapes.}
We use synthetic utility landscapes to model defender-side utilities over attack subsets.
These experiments do not synthesize jailbreak attacks; instead, each task defines a bounded utility oracle \(v(S)\in[0,1]\) over a fixed attack universe \(\mathcal{A}\).
Each task contains \(n=20\) candidate attacks, allowing exact enumeration of all \(2^{20}\) subset utilities and exact reference AttackSHAP values.
We instantiate 90 utility landscapes from scalar, vector, and random generation policies.
Across policies, the construction is designed to capture patterns we expect in real defense utilities, including diminishing gains, redundant attacks, and complementary attack combinations.
Full construction details and an example utility-landscape visualization are provided by Figure~\ref{fig:utility_landscape} in Appendix.

\paragraph{Real LLM safety settings.}
Following the workflow in Figure~\ref{fig:placeholder}, each real setting is specified by a target model, a defense algorithm, defense data constructed from selected attacks, and held-out test attacks, test jailbreak/over-refusal datasets used to compute defender-centric utility.
We evaluate three settings with different target models and complementary goals.
First, to test whether attacker-centric success predicts defender-centric utility, we use Llama-3-8B with in-context defense~\cite{wei2023jailbreak}, construct defense demonstrations from JailbreakBench~\cite{chao2024jailbreakbench} attacks, and compare no-defense ASR rankings with singleton defense utility evaluated on JailbreakBench and SorryBench~\cite{xie2024sorry}.
Second, to evaluate utility learning and AttackSHAP estimation, we use Qwen2.5-7B with adversarial training~\cite{mazeika2024harmbench}, where selected JailbreakBench attacks generate safety-training data and \(v(S)\) is measured by the resulting robustness utility.
Third, to evaluate budgeted subset selection, we use Ministral-3-8B with in-context defense; selected attacks form the defense context on JailbreakBench, while utility is computed from held-out JailbreakBench/SorryBench robustness together with over-refusal~\cite{rottger2024xstest}.
Across settings, the attack universe \(\mathcal{A}\) contains 10 attack categories instantiated on JailbreakBench and SorryBench, covering 
template-based~\cite{wei2023jailbreak,shen2024anything,yoosuf2025structtransform,zhu2024advprefix}, 
strategy-based~\cite{andriushchenko2025does,huang2026obscure,luo2026simple}, and 
learning-based~\cite{zou2023universal,zhou2025don,zhu2024autodan} jailbreaks.
The two defense algorithms, adversarial training and in-context defense, instantiate different ways in which an attack subset \(S\) changes the defended system.
For subset selection, we evaluate budgets \(k\in\{2,4\}\).
Full implementation details, attack taxonomy, and hyperparameters are provided in Appendix.

\paragraph{Compared methods.}
Table~\ref{tab:compared_methods} summarizes the compared methods. 
We separate methods that produce attribution scores from methods that directly return a size-\(k\) subset. 
Attribution methods can be converted into subset selectors by ranking attacks and taking the top \(k\), 
while selection methods optimize the subset-level decision more directly.
\begin{table*}[t]
\centering
\small
\setlength{\tabcolsep}{4pt}
\vspace{-2em}
\resizebox{0.85\textwidth}{!}{
\begin{tabular}{llp{9cm}}
\toprule
\textbf{Method} & \textbf{Utility Budget} & \textbf{Description} \\
\midrule

\multicolumn{3}{l}{\footnotesize \textcolor{gray}{\textit{Attribution methods}}} \\

\textsc{ASR} & \(0\) & Attacker-centric ranking by attack success rate \\
\textsc{AOO} & \(n+1\) & Empty-context marginal utility gain \\
\textsc{LOO} & \(n+1\) & Full-context marginal utility loss \\
\textsc{Standard SHAP}\(_v\) & \(2^n\) & Reference Shapley attribution from true utilities \\
\textsc{AttackSHAP}\(_{\hat v}\) & \(M\) & Shapley attribution from learned surrogate utilities \\
\textsc{SHAP Baselines} & \(M\) & Standard Shapley value estimation baselines, including DataSHAP \cite{ghorbani2019data}, SVA \cite{zhang2023efficient}, SamplingSHAP@K \cite{kariyappa2024shap} \\

\addlinespace[3pt]
\multicolumn{3}{l}{\footnotesize \textcolor{gray}{\textit{Subset selection methods}}} \\
\textsc{Random} & \(0\) & Random size-\(k\) subset \\
\textsc{TopK}(\(\cdot\)) & score-dependent & Select top \(k\) attacks by an attribution score listed above \\
\textsc{A-MESS-Greedy} & \(nk-k(k-1)/2\) & Greedy selection using true marginal utility \\
\textsc{A-MESS-Surrogate} & \(M\) & Search over the learned utility landscape \\
\textsc{OracleSearch}\(_v\) & \(\binom{n}{k}\) & Exhaustive true-utility search; not feasible and acts as upper bound only \\
\bottomrule
\end{tabular}
}
\caption{Compared methods. Utility budget denotes the number of true subset-utility evaluations \(v(S)\), excluding attacker-centric quantities such as ASR. \(M\) is the utility budget specified by each concrete method.}
\label{tab:compared_methods}
\end{table*}

\begin{table*}[t]
\centering
\small
\resizebox{0.85\textwidth}{!}{
\begin{tabular}{lrrrrr}
\toprule
Attack & ASR Rank & No-defense ASR & Utility Rank & \(v_\theta(\{a\})\) & \(\Delta\)Rank \\
\midrule
\texttt{Optimized Suffix}     & 1  & 0.620 & 7 & 0.846 & -6 \\
\texttt{Refusal Suppression}  & 2  & 0.270 & 1 & 0.904 & +1 \\
\texttt{Roleplay}             & 9  & 0.000 & 2 & 0.890 & +7 \\
\midrule
\multicolumn{6}{l}{Spearman $\rho=0.215$, Kendall $\tau=0.190$.} \\
\bottomrule
\end{tabular}
}
\caption{
Representative ranking shifts between attacker-centric success and defender-centric utility.
ASR Rank is computed from no-defense attack success rate, while Utility Rank is computed from singleton in-context-defense utility. 
We define \(\Delta\text{Rank} = \text{ASR Rank} - \text{Utility Rank}\), so positive values indicate that an attack ranks higher under defender-side utility than under attack success.
}
\label{tab:asr_utility_rank_examples}
\vspace{-2em}
\end{table*}

\subsection{From Attack Success to Defense Utility}
We first examine whether attacker-centric success is aligned with defender-centric utility in a real LLM safety setting. 
For each attack category \(a\in\mathcal{A}\), we compute its no-defense ASR on JailbreakBench~\cite{chao2024jailbreakbench} to obtain the attacker-centric score. 
We then compute its singleton defense utility \(v_\theta(\{a\})\) under the in-context defense setting, where the defense is constructed using only that attack category and evaluated by the corresponding defender-side safety utility. 
If ASR were a reliable proxy for defender-centric utility, the no-defense ASR ranking and the singleton utility ranking should be largely consistent.

\begin{figure}[H]
    \centering
    \vspace{.5em}
    \hspace*{-0.13\columnwidth} 
    \includegraphics[width=.6\textwidth]{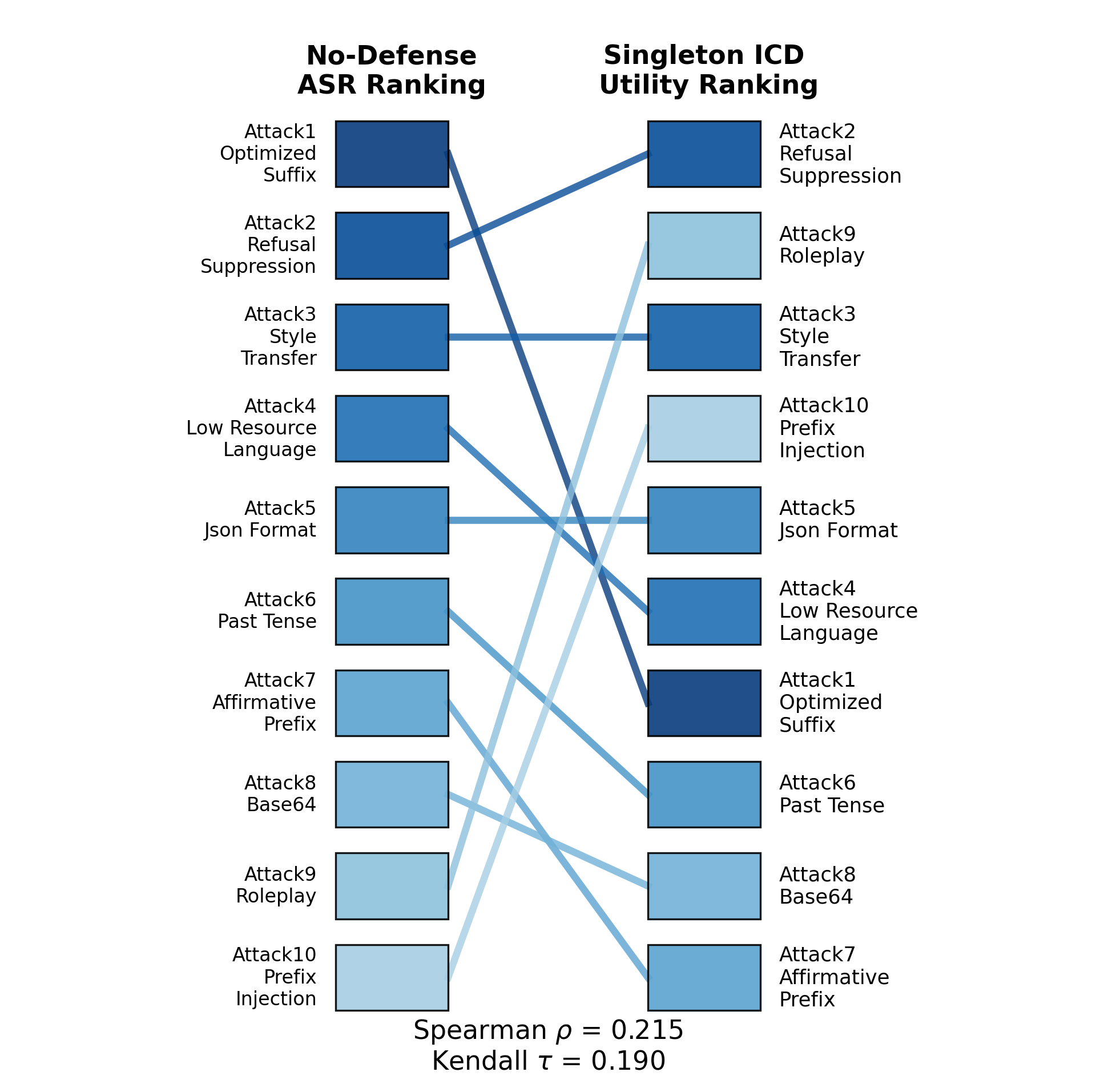}
    \caption{
    Attack ASR ranking differs from defender-centric utility ranking.
    Attack categories are ordered by no-defense ASR on the left and by singleton in-context-defense utility \(v_\theta(\{a\})\) on the right.
    Each line connects the same attack across two rankings.
    The low rank correlation, Spearman \(\rho = 0.215\) and Kendall \(\tau = 0.190\), shows that attacker-centric success is an incomplete proxy for defender-centric utility in this setting.
    }
    \vspace{-1em}
    \label{fig:asr_vs_utility_rank}
\end{figure}

However, Figure~\ref{fig:asr_vs_utility_rank} shows that this is not the case. 
The two rankings are weakly correlated, with Spearman \(\rho = 0.215\) and Kendall \(\tau = 0.190\). 
Several attacks move substantially across the two rankings, as demonstrated in Table~\ref{tab:asr_utility_rank_examples}.
For example, \textsc{Roleplay} has zero no-defense ASR, yet becomes the second highest-ranked attack by singleton defense utility. 
In contrast, \textsc{Optimized Suffix} has the highest no-defense ASR, but drops to seventh place under defender-side utility. 
These results suggest that attacks that most effectively break models are not necessarily those that most effectively improve defense utility. 
Therefore, attacker-centric success is an incomplete criterion for identifying which attacks matter for safety alignment.


\begin{figure*}[t]
    \vspace{-1.5em}
    \centering
    \begin{subfigure}[t]{0.66\textwidth}
        \centering
        \includegraphics[width=\linewidth]{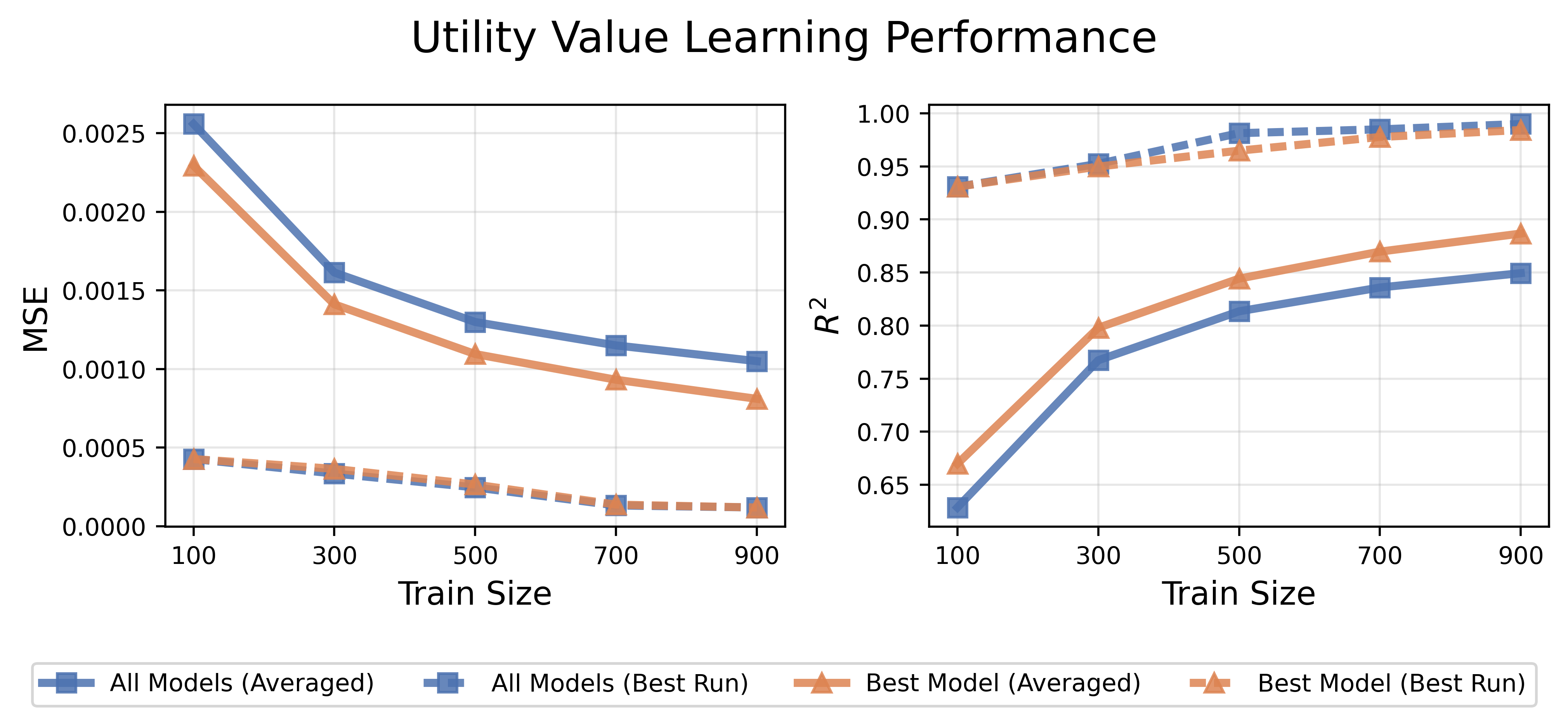}
        \caption{
        Surrogate utility learning performance. We evaluate how accurately the learned surrogate utility model approximates the ground-truth subset utility under different numbers of sampled utility evaluations.
        The surrogate model increasingly approximates the underlying subset utility landscape as more sampled subset evaluations are provided.
        $R^2$ achieves more than 0.9 for the best performing surrogate models.
        }
        \label{fig:synthetic_v_learning}
    \end{subfigure}
    \hfill
    \begin{subfigure}[t]{0.3\textwidth}
        \centering
        \includegraphics[width=\linewidth]{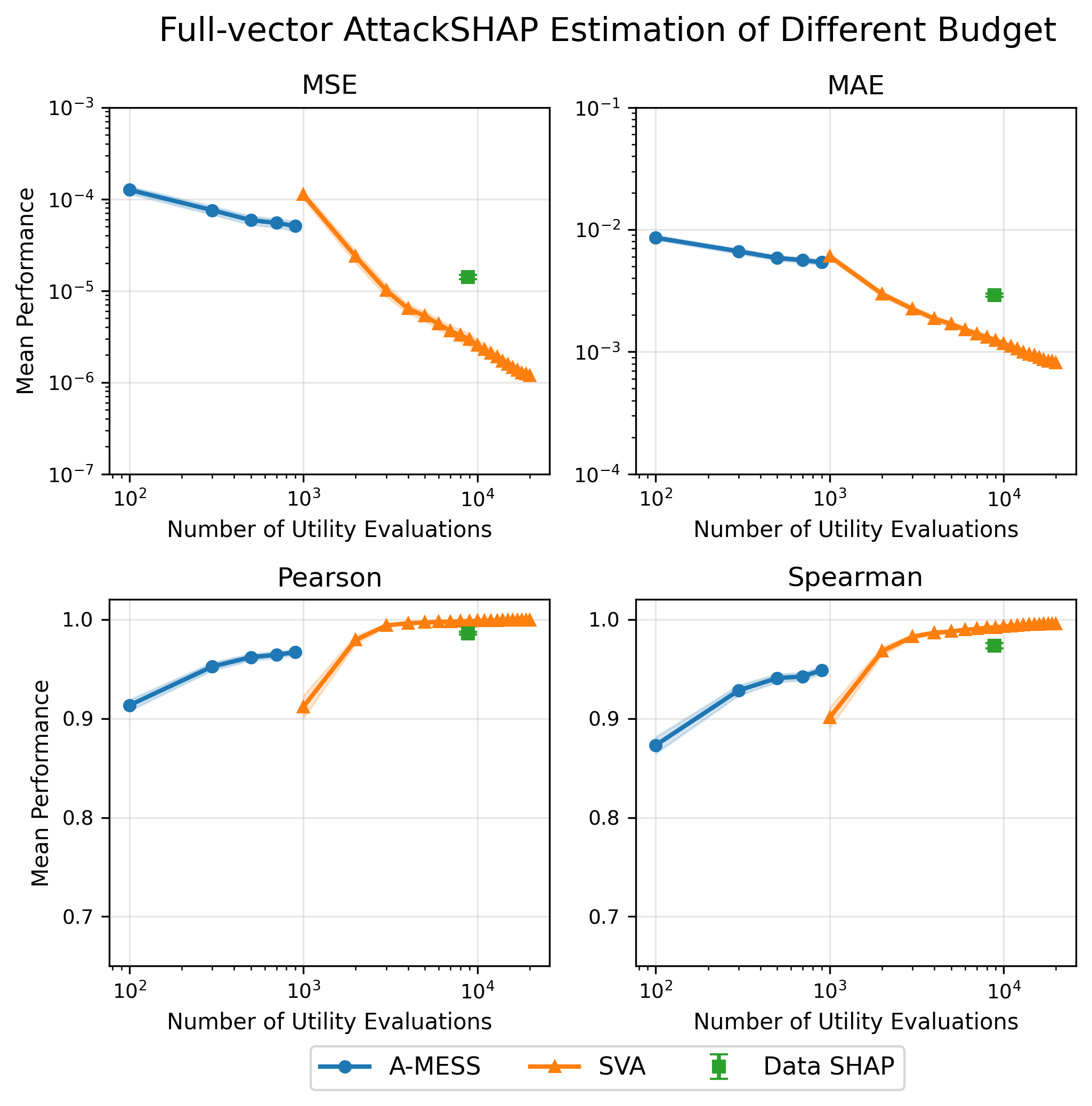}
        \caption{
        Full-vector AttackSHAP estimation. We compare AttackSHAP estimation quality under different utility budgets using error-based and correlation-based metrics.
        }
        \label{fig:synthetic_full_vector_shap}
    \end{subfigure}

    \vspace{0.2em}

    \begin{subfigure}[t]{\textwidth}
        \centering
        \includegraphics[width=\linewidth]{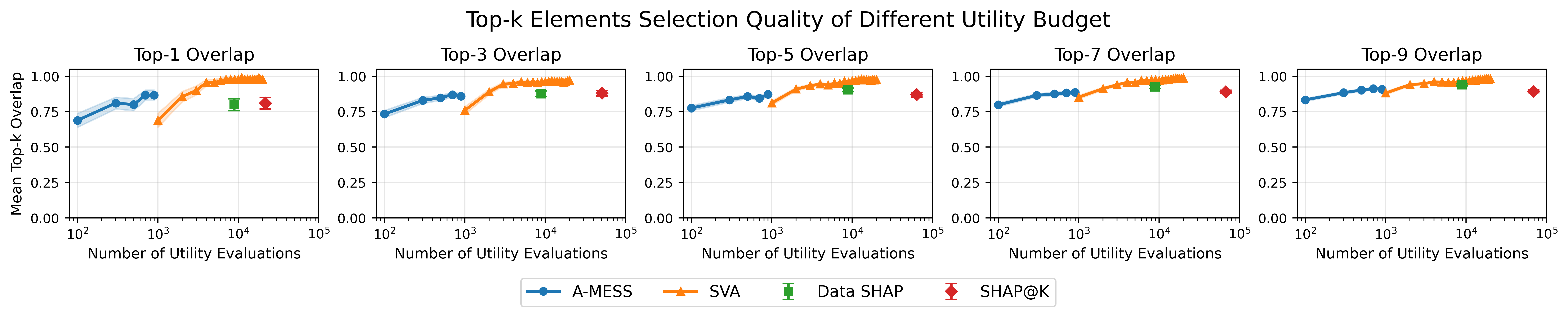}
        \caption{
        Top-\(k\) attack attribution quality. We evaluate whether estimated attribution scores recover the top-\(k\) reference attacks.
        }
        \label{fig:synthetic_topk_shap}
    \end{subfigure}
    \vspace{-.5em}
    \caption{
    Synthetic evaluation of \textsc{A-MESS-SURROGATE}.
    Panel~\ref{fig:synthetic_v_learning} evaluates surrogate utility prediction under increasing utility-query budgets.
    Panel~\ref{fig:synthetic_full_vector_shap} compares surrogate-based AttackSHAP estimates with reference full-vector AttackSHAP values.
    Panel~\ref{fig:synthetic_topk_shap} measures top-\(k\) overlap with the attacks identified by reference AttackSHAP.
    }
    \vspace{-1em}
    \label{fig:shap_estimation}
\end{figure*}

\begin{table}[t]
\centering
\small
\setlength{\tabcolsep}{4pt}
\resizebox{0.46\textwidth}{!}{
\begin{tabular}{lcccc}
\toprule
$n_{\mathrm{train}}$:$n_{\mathrm{test}}$ & MAE $\downarrow$ & $R^2$ $\uparrow$ & Pearson $\uparrow$ & Spearman $\uparrow$ \\
\midrule
50:250 & 0.078{\scriptsize $\pm$ 0.005} & 0.779{\scriptsize $\pm$ 0.031} & 0.888{\scriptsize $\pm$ 0.017} & 0.893{\scriptsize $\pm$ 0.019} \\
100:200 & 0.071{\scriptsize $\pm$ 0.003} & 0.817{\scriptsize $\pm$ 0.018} & 0.908{\scriptsize $\pm$ 0.008} & 0.914{\scriptsize $\pm$ 0.009} \\
150:150 & 0.066{\scriptsize $\pm$ 0.004} & 0.828{\scriptsize $\pm$ 0.026} & 0.912{\scriptsize $\pm$ 0.015} & 0.917{\scriptsize $\pm$ 0.016} \\
200:100 & 0.067{\scriptsize $\pm$ 0.006} & 0.825{\scriptsize $\pm$ 0.038} & 0.912{\scriptsize $\pm$ 0.021} & 0.917{\scriptsize $\pm$ 0.020} \\
\bottomrule
\end{tabular}
}
\vspace{-.5em}
\caption{Real-LLM subset utility prediction. We train the \textsc{A-MESS-Surrogate} on observed subset utilities and evaluate held-out prediction of $v(S)$. Metrics are mean $\pm$ std across random splits.}
\label{tab:task1_real_v_prediction}
\vspace{-1em}
\end{table}


\begin{table}[H]
\centering
\small
\resizebox{0.46\textwidth}{!}{
\begin{tabular}{lccccc}
\toprule
\textbf{Method}
& \(\boldsymbol{k=1}\)
& \(\boldsymbol{k=3}\)
& \(\boldsymbol{k=5}\)
& \(\boldsymbol{k=7}\)
& \(\boldsymbol{k=9}\) \\
\midrule

\multicolumn{6}{l}{\footnotesize \textcolor{gray}{\textit{Attribution Baselines}}} \\
\(\textsc{TopK-AOO}_{v}\)
& \textbf{100.00} & 53.12 & 46.66 & 37.76 & 31.52 \\

\(\textsc{TopK-LOO}_{v}\)
& 33.99 & 36.48 & 44.79 & 57.33 & 63.80 \\

\(\textsc{TopK-}\phi_{v}\)
& 65.39 & 61.10 & 67.03 & 68.51 & 69.28 \\

\(\textsc{TopK-}\hat{\phi}\)
& 59.66 & 57.37 & 62.94 & 65.01 & 71.07 \\

\midrule
\addlinespace[3pt]
\multicolumn{6}{l}{\footnotesize \textcolor{gray}{\textit{A-MESS Solvers}}} \\

\(\textsc{A-MESS-Greedy}\)
& \textbf{100.00} & \textbf{85.47} & \textbf{83.36} & \textbf{83.20} & 82.69 \\

\(\textsc{A-MESS-Surrogate}\)
& 59.78 & 73.18 & 74.48 & 79.79 & \textbf{84.52} \\

\bottomrule
\end{tabular}
}
\vspace{-.5em}
\caption{
Synthetic subset selection results. We report Normalized Gain \(NG_k\) (\%) averaged over 90 utility landscapes, where higher values indicate closer recovery of the oracle subset utility.
}
\label{tab:synthetic_ng}
\vspace{-1.5em}
\end{table}

\subsection{Budgeted AttackSHAP Estimation}

This subsection evaluates A-MESS-Surrogate at two levels. 
On synthetic landscapes, where the full utility landscape and reference AttackSHAP is available, we test whether the surrogate recovers both the utility function and the resulting AttackSHAP values.
Existing Shapley estimators are compared.
On real LLM utilities, where exhaustive evaluation is infeasible, we test held-out prediction of \(v_\theta(S)\) as the observable proxy.

We first evaluate whether \textsc{A-MESS-Surrogate} can estimate AttackSHAP with limited utility queries.
On synthetic landscapes, the full \(2^{20}\) utility table provides oracle AttackSHAP references.
Figure \ref{fig:shap_estimation} shows that the learned utility model reaches high \begin{small}\(R^2\)\end{small} in the high-budget regime.
More importantly, the AttackSHAP estimates match or outperform strong existing Shapley estimators on both full-vector $\phi$ prediction and top-\(k\) coverage, while requiring only hundreds of true utility queries rather than thousands or tens of thousands.
A-MESS reaches a strong accuracy efficiency tradeoff for standard Shapley estimation.


We further evaluate the same surrogate on real Qwen2.5-7B adversarial-training utilities.
Since the full real utility landscape cannot be exhaustively observed, we test held-out prediction of \(v(S)\), reporting mean and standard deviation over five random train/test splits for each training-set size.
Table~\ref{tab:task1_real_v_prediction} shows that the surrogate already obtains strong utility prediction performance with 150 training subsets, 
and the improvement from 150 to 200 training subsets is small,
suggesting convergence.
This indicates that, in this real LLM setting, roughly 200 observed subset utilities are sufficient for A-MESS to learn a practical approximation of the defender-centric utility landscape.

\subsection{Minimal Effective Subset Selection}
\label{sec:subset-selection}
\begin{table*}[t]
\centering
\scriptsize
\setlength{\tabcolsep}{3pt}
\renewcommand{\arraystretch}{1.08}

\begin{tabularx}{\textwidth}{
@{} l l c c
@{\hspace{1.2em}}
l l c c @{}
}
\toprule
Method & Selected subset & \makecell[c]{Utility$\uparrow$} & \makecell[c]{$\Delta U$$\uparrow$}
&
Method & Selected subset & \makecell[c]{Utility$\uparrow$} & \makecell[c]{$\Delta U$$\uparrow$}
\\
\midrule
\textsc{No Defense}
& --
& 0.1960
& 0.0000
&
\textsc{No Defense}
& --
& 0.1960
& 0.0000
\\
\midrule
\multicolumn{4}{@{}l@{}}{\textit{\textcolor{gray}{Baselines with budget $k=2$}}}
&
\multicolumn{4}{@{}l@{}}{\textit{\textcolor{gray}{Baselines with budget $k=4$}}}
\\
\textsc{Random}
& Averaged over 5 random subsets
& 0.6183
& 0.4223
&
\textsc{Random}
& Averaged over 5 random subsets
& 0.5342
& 0.3382
\\

\textsc{TopK-ASR}
& Attack$_1$, Attack$_8$
& 0.5716
& 0.3756
&
\textsc{TopK-ASR}
& Attack$_1$, Attack$_2$, Attack$_3$, Attack$_8$
& 0.8613
& 0.6653
\\

\textsc{TopK-AOO}
& Attack$_2$, Attack$_4$
& 0.5379
& 0.3419
&
\textsc{TopK-AOO}
& Attack$_0$, Attack$_2$, Attack$_4$, Attack$_5$
& 0.5448
& 0.3488
\\

\textsc{TopK-LOO}
& Attack$_2$, Attack$_8$
& 0.6447
& 0.4487
&
\textsc{TopK-LOO}
& Attack$_2$, Attack$_7$, Attack$_8$, Attack$_9$
& 0.5616
& 0.3656
\\

\midrule
\multicolumn{4}{@{}l@{}}{\textit{\textcolor{gray}{A-MESS Solvers with budget $k=2$}}}
&
\multicolumn{4}{@{}l@{}}{\textit{\textcolor{gray}{A-MESS Solvers with budget $k=4$}}}
\\

\textsc{A-MESS-Greedy}
& Attack$_1$, Attack$_2$
& \textbf{0.8539}
& \textbf{0.6579}
&
\textsc{A-MESS-Greedy}
& Attack$_1$, Attack$_2$, Attack$_3$, Attack$_6$
& \textbf{0.9397}
& \textbf{0.7437}
\\

\textsc{A-MESS-Surrogate}
& Attack$_1$, Attack$_2$
& \textbf{0.8539}
& \textbf{0.6579}
&
\textsc{A-MESS-Surrogate}
& Attack$_1$, Attack$_2$, Attack$_3$, Attack$_6$
& \textbf{0.9397}
& \textbf{0.7437}
\\
\bottomrule
\end{tabularx}

\vspace{-.5em}
\caption{
Real in-context-defense subset selection results on Ministral-3.
Utility is defined as
\(U=0.8*\mathrm{Score_{ASR}}+0.2*\mathrm{Score_{ORR}}\),
where \(\mathrm{Score}_{\mathrm{ORR}}=1-\mathrm{ORR}\) denotes over-refusal rate on XSTest.
\(\Delta U\) denotes the absolute utility gain over the no-defense baseline.
Both \textsc{A-MESS} solvers select the highest-utility subset under both \(k\) budget.
}
\label{tab:real_subset_selection}
\vspace{-1em}
\end{table*}


We finally evaluate whether attack attribution directly yields the best subset, and whether A-MESS can improve subset-level utility. 
Table~\ref{tab:synthetic_ng} reports Normalized Gain on 90 synthetic utility landscapes. 
Attribution-based top-\(k\) selection improves over random selection, but remains clearly below direct subset optimization. 
Among attribution baselines, \textsc{TopK}-\(\phi_v\) and \textsc{TopK}-\(\hat{\phi}\) are the strongest, reaching roughly 60--71\% Normalized Gain for nontrivial budgets. 
In contrast, \textsc{A-MESS-Greedy} recovers at least 82.69\% of the oracle improvement gap for \(k>1\), while \textsc{A-MESS-Surrogate} improves as the budget grows and reaches 84.52\% at \(k=9\). 
These results show that ranking attacks by individual contribution is not sufficient, 
and explicitly optimizing the subset can recover more oracle gain.


Table~\ref{tab:real_subset_selection} evaluates the same question in the real ICD setting on Ministral-3. 
A-MESS selects the highest-utility subset under both evaluated budgets. 
Both solvers choose $\{1,2\}$ for \(k=2\), reaching top utility 0.8539, and $\{1,2,3,6\}$ for \(k=4\), reaching top utility 0.9397. 
This trend is consistent with the synthetic results, that
high-utility attack subsets are not necessarily recovered by attacker-centric metrics or individual attribution.

\section{Discussion}

The main contribution of this work is not merely a numerical improvement, but a shift in how jailbreak attacks should be interpreted in safety research. 
Much of jailbreak research has been driven by an arms race.
However, a first-principles view of doing research should start from the very final goal~\cite{lu2025discovery}.
Applied to jailbreak research, this view suggests that attacks should be valued not only by how often they break an undefended model, but by how much they improve downstream safety when used as red-teaming resources. 
Our results make this distinction concrete: ASR and singleton defense utility are weakly aligned in our real setting, and attacks with high individual attribution do not necessarily form the best subset.
This reframes jailbreak attacks as defense resources whose value depends on the downstream safety objective.

This perspective also motivates the black-box, setting-agnostic design of A-MESS. Safety settings may be largely heterogeneous and proprietary, and many production pipelines are only partially observable. 
Rather than seeking a universally best subset, as a setting-agnostic framework A-MESS requires only utility observations \(v_\theta(S)\), allowing practitioners to instantiate the framework under their own safety objectives. This does not solve the broader transparency gap, but it provides a practical interface for connecting jailbreak evaluation to measurable safety improvement.

\section{Related Work}

\paragraph{Jailbreak attacks and safety defenses.}
Prior work has proposed diverse jailbreak attacks against aligned LLMs, including 
template-based~\cite{wei2023jailbreak,shen2024anything,yoosuf2025structtransform,zhu2024advprefix}, 
strategy-based~\cite{andriushchenko2025does,huang2026obscure,luo2026simple}, and 
learning-based attacks~\cite{zou2023universal,zhou2025don,zhu2024autodan}.
These attacks are typically evaluated by attacker-centric metrics such as ASR, transferability, universality, and cost. 
In parallel, LLM safety work has studied refusal training, in-context defenses, guardrails, and over-refusal-aware evaluation~\cite{yu2024robust,mo2024fight,fu2025short,rottger2024xstest,zhang2025safety,cui2025orbench,pan2025understanding}. 
These defenses show that attack data can improve safety, but rarely ask which attacks are most useful for downstream safety alignment. 
A-MESS bridges this gap by treating jailbreak attacks as red-teaming resources and evaluating them through setting-specific defender-side utility.

\paragraph{Game-theoretic valuation and subset selection.}
Shapley values have been widely used to attribute machine-learning utility to data points, features, or training examples \cite{shapley1953value,dubey1975uniqueness,ghorbani2019data},
with later work studying more efficient Shapley approximation and top-\(k\) identification
\cite{zhang2023efficient,kariyappa2024shap}.
A-MESS differs in both the object being valued and the decision it supports: the players are jailbreak attacks, the coalition value is downstream defender-side utility under a safety pipeline, and attribution is paired with budgeted subset selection rather than treated as the final goal.
\section{Conclusion}
We presented A-MESS, a defender-centric framework for attributing and selecting jailbreak attacks by their contribution to downstream safety utility.
Across controlled utility landscapes and real settings, our results show that 
high-ASR attacks need not yield high defender-side utility,
and that high-contribution attacks do not necessarily form the best subset.
These findings suggest that jailbreak evaluation should go beyond attack success alone and ask which attacks help build safer models.

\clearpage
\section*{Limitations}

A-MESS assumes that the defender-side utility of an attack subset is meaningful, which requires the defense procedure to be coupled with the selected attacks.
This includes settings such as adversarial training, in-context defense, or other pipelines where different attack subsets can lead to different defended systems.
The framework is less suitable for defenses whose behavior is independent of the selected attack set.
In such cases~\cite{xie2023defending}, \(v_\theta(S)\) may not meaningfully vary with \(S\), making attack attribution or subset selection ill-posed.

LLM safety alignment also includes downstream concerns beyond jailbreak robustness, such as backdoor robustness~\cite{yang2024watch}, watermarking~\cite{dathathri2024scalable}, hallucination~\cite{ji2023towards}, and broader reliability evaluation~\cite{zhou2025beyond}.
Since A-MESS only requires a setting-specific subset utility \(v_\theta(S)\), the framework is in principle applicable to these settings as well.
Due to space and computation constraints, however, this paper focuses on jailbreak attacks as the primary downstream task, leaving broader safety-alignment instantiations to future work.


\bibliography{my_ref}

\clearpage
\appendix
\section{Appendix Overview}

This appendix provides additional details for the experimental settings used in the main paper.
Appendix~\ref{app:synthetic_details} describes the construction of synthetic utility landscapes and the utility-landscape visualization.
Appendix~\ref{app:synthetic_raw_results} reports raw synthetic subset utilities corresponding to the Normalized Gain results in the main paper.
Appendix~\ref{app:real_llm_details} summarizes the real LLM safety settings.
Appendix~\ref{app:attack_taxonomy} lists the attack categories used in the real LLM experiments.
Appendix~\ref{app:baseline_details} gives implementation details for baselines and A-MESS solvers.

\section{Synthetic Utility Landscapes}
\label{app:synthetic_details}

\paragraph{Purpose.}
The synthetic experiments are designed to evaluate A-MESS under controlled defender-side utility functions where the full subset utility table is observable.
They do not synthesize jailbreak prompts.
Instead, each synthetic task defines a set function \(v(S)\) over subsets \(S\subseteq\mathcal{A}\), which allows us to compute exact AttackSHAP values, oracle size-\(k\) subsets, and Normalized Gain.

\paragraph{Task construction.}
Each synthetic task contains \(n=20\) candidate attacks, so the full utility landscape contains \(2^{20}\) subset values.
We instantiate 90 landscapes using three generation policies: scalar, vector, and random.
Each policy contributes 30 landscapes with different random seeds.
All utilities are min-max normalized within each landscape to the range \([0,1]\).

\paragraph{Generation policies.}
The scalar policy assigns each attack a latent scalar strength and aggregates the selected strengths.
The vector policy assigns each attack a vector in a latent coverage space and scores the subset by the coverage induced by the selected vectors.
The random policy starts from per-attack base potentials and adds deterministic subset-level variation.
Across all policies, the latent score is further shaped by shared mechanisms that mimic patterns expected in real defense utilities: diminishing gains as subsets grow, a weak diversity bonus for covering more synthetic attack families, and pairwise interactions that can model either redundancy or complementarity between attacks.

\paragraph{Exact references.}
For each landscape, we enumerate all \(2^{20}\) subset utilities and compute exact AttackSHAP values from the full table.
These exact values serve as references for evaluating surrogate utility learning, full-vector AttackSHAP estimation, top-\(k\) attribution recovery, and budgeted subset selection.
For surrogate learning, sampled subset utilities are split into training and validation subsets, and held-out utility prediction is evaluated on unqueried subsets from the same landscape.

\begin{figure*}[t]
    \centering
    \includegraphics[width=0.78\linewidth]{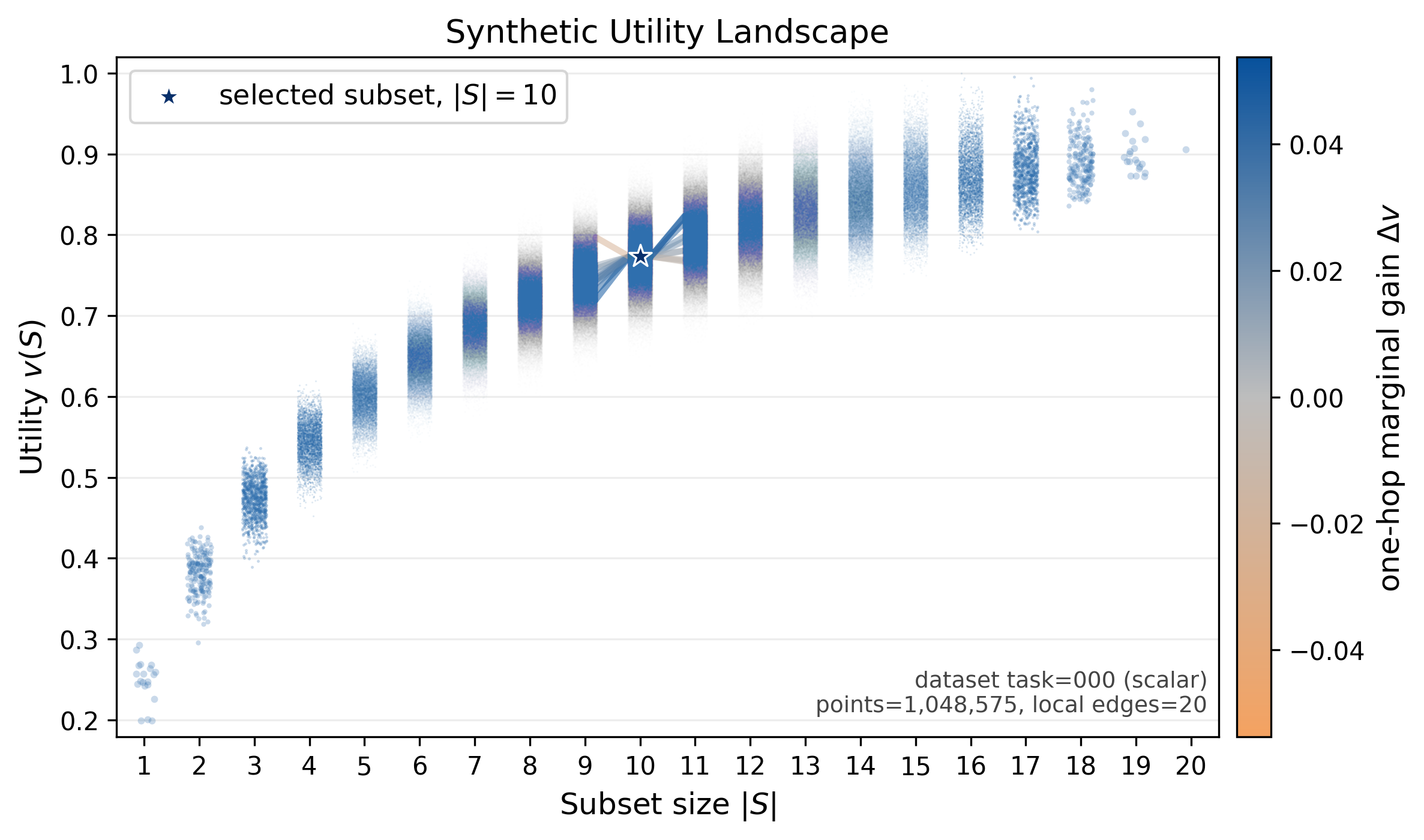}
    \caption{
    Example synthetic utility landscape.
    Each point corresponds to one attack subset, grouped by subset size \(|S|\) on the x-axis and plotted by utility \(v(S)\) on the y-axis.
    The highlighted star marks one selected subset, and local edges show one-hop neighboring subsets obtained by adding or removing one attack.
    Edge color indicates the one-hop marginal utility change \(\Delta v\).
    }
    \label{fig:utility_landscape}
\end{figure*}

\paragraph{Utility-landscape visualization.}
Figure~\ref{fig:utility_landscape} visualizes one representative synthetic landscape.
It shows how subset utility varies across subset sizes and illustrates that neighboring subsets can have different marginal gains even when their sizes are similar.
This provides an intuitive view of why subset-level utility cannot always be recovered by ranking attacks independently.

\section{Additional Synthetic Subset-Selection Results}
\label{app:synthetic_raw_results}

The main text reports Normalized Gain for synthetic subset selection because it normalizes each utility landscape by its own random-baseline and oracle gap before averaging.
For completeness, Table~\ref{tab:subset_raw_utility} reports the corresponding raw subset utility \(v_\theta(S_k(m))\), averaged over the same 90 minmax utility landscapes used in Table~\ref{tab:synthetic_ng}.
Because Normalized Gain is averaged per landscape, it is not obtained by normalizing the marginal means in Table~\ref{tab:subset_raw_utility}.
The raw utilities show the same qualitative pattern as Normalized Gain: direct subset optimization improves over attribution-only top-\(k\) selection, while \textsc{OracleSearch}\(_v\) remains the upper bound.

\begin{table*}[t]
\centering
\small
\resizebox{0.65\linewidth}{!}{
\begin{tabular}{lccccc}
\toprule
\textbf{Method}
& \(\boldsymbol{k=1}\)
& \(\boldsymbol{k=3}\)
& \(\boldsymbol{k=5}\)
& \(\boldsymbol{k=7}\)
& \(\boldsymbol{k=9}\) \\
\midrule

\multicolumn{6}{l}{\footnotesize \textcolor{gray}{\textit{Natural Baseline}}} \\
\(\textsc{Random}\)
& 0.2941 & 0.4527 & 0.5479 & 0.6171 & 0.6663 \\

\addlinespace[3pt]
\multicolumn{6}{l}{\footnotesize \textcolor{gray}{\textit{Attribution Baselines}}} \\
\(\textsc{TopK-AOO}_{v}\)
& 0.4456 & 0.5580 & 0.6471 & 0.6980 & 0.7329 \\

\(\textsc{TopK-LOO}_{v}\)
& 0.3497 & 0.5208 & 0.6344 & 0.7344 & 0.8055 \\

\(\textsc{TopK-}\phi_{v}\)
& 0.4018 & 0.5641 & 0.6814 & 0.7592 & 0.8158 \\

\(\textsc{TopK-}\hat{\phi}\)
& 0.3828 & 0.5577 & 0.6682 & 0.7506 & 0.8203 \\

\addlinespace[3pt]
\multicolumn{6}{l}{\footnotesize \textcolor{gray}{\textit{A-MESS Solvers}}} \\
\(\textsc{A-MESS-Greedy}\)
& 0.4456 & 0.6119 & 0.7143 & 0.7897 & 0.8449 \\

\(\textsc{A-MESS-Surrogate}\)
& 0.3791 & 0.5833 & 0.6958 & 0.7841 & 0.8527 \\

\addlinespace[3pt]
\multicolumn{6}{l}{\footnotesize \textcolor{gray}{\textit{Oracle Upper Bound}}} \\
\(\textsc{OracleSearch}_{v}\)
& 0.4456 & 0.6365 & 0.7450 & 0.8253 & 0.8851 \\

\bottomrule
\end{tabular}
}
\caption{
Raw subset utility averaged over the same 90 synthetic utility landscapes.
For each method \(m\), \(S_k(m)\) denotes the selected size-\(k\) subset, and each entry reports \(v_\theta(S_k(m))\).
These marginal means are reported for interpretability, and the Normalized Gain in Table~\ref{tab:synthetic_ng} is computed per landscape before averaging.
}
\label{tab:subset_raw_utility}
\end{table*}

\section{Real LLM Safety Settings}
\label{app:real_llm_details}

Table~\ref{tab:real_settings_appendix} summarizes the three real LLM safety settings used in the main experiments.

\begin{table*}[t]
\centering
\small
\setlength{\tabcolsep}{3pt}
\renewcommand{\arraystretch}{1.12}
\caption{Real LLM safety settings used in the main experiments.}
\label{tab:real_settings_appendix}
\begin{tabularx}{\textwidth}{p{0.19\textwidth}p{0.13\textwidth}p{0.1\textwidth}p{0.25\textwidth}X}
\toprule
\textbf{Setting} & \textbf{Target model} & \textbf{Defense pipeline} & \textbf{Defense / test data} & \textbf{Utility and role in paper} \\
\midrule
ASR--utility alignment
& Llama-3-8B
& In-context defense
& Defense demonstrations are constructed from singleton JailbreakBench attack categories; held-out JailbreakBench and SorryBench attacks are used for safety evaluation.
& Compares no-defense ASR ranking with singleton defender-side utility ranking to test whether attacker-centric success is a reliable proxy. \\
\addlinespace
AttackSHAP estimation
& Qwen2.5-7B
& Adversarial training
& Selected JailbreakBench attacks generate safety-training data; held-out JailbreakBench attacks are used to measure post-defense robustness.
& Provides real subset-utility observations for learning \(\hat v_\theta\); held-out prediction of \(v_\theta(S)\) evaluates whether the surrogate captures the real utility landscape. \\
\addlinespace
Subset selection
& Ministral-3-8B
& In-context defense
& Selected JailbreakBench attacks form the defense context; held-out JailbreakBench and SorryBench attacks evaluate jailbreak robustness, and XSTest evaluates over-refusal.
& Evaluates size-\(k\) subset selection with \(k\in\{2,4\}\), using \(U=0.8\,\mathrm{Score}_{\mathrm{ASR}}+0.2\,\mathrm{Score}_{\mathrm{ORR}}\). \\
\bottomrule
\end{tabularx}
\end{table*}

\paragraph{Defense data versus test data.}
In all real LLM settings, defense data and test data are separated.
Defense data are generated from the selected attack subset \(S\) and are used either to construct in-context demonstrations or to create adversarial-training examples.
Test data are held out and are used only to compute the final utility \(v_\theta(S)\).
For in-context defense, selected attacks change the defended prompt at inference time.
For adversarial training, selected attacks change the fine-tuning data and therefore the resulting defended model.

\paragraph{Utility definitions.}
For robustness-only settings, utility is based on the post-defense reduction of jailbreak ASR.
For the subset-selection setting, \(\mathrm{Score}_{\mathrm{ASR}}\) is the relative balanced-ASR gain over the no-defense baseline, clipped to \([0,1]\), and \(\mathrm{Score}_{\mathrm{ORR}}=1-\mathrm{ORR}\), where ORR is the over-refusal rate on XSTest.
This utility rewards jailbreak robustness while penalizing defenses that over-refuse benign requests.

\section{Attack Taxonomy}
\label{app:attack_taxonomy}

Table~\ref{tab:attack_taxonomy_appendix} lists the ten attack categories used across the real LLM experiments.

\begin{table*}[t]
\centering
\small
\setlength{\tabcolsep}{5pt}
\renewcommand{\arraystretch}{1.08}
\caption{Attack categories used in the real LLM experiments.}
\label{tab:attack_taxonomy_appendix}
\begin{tabularx}{\textwidth}{p{0.5\textwidth}X}
\toprule
\textbf{Attack name} & \textbf{Brief description} \\
\midrule
Base64~\cite{chu2025jailbreakradar} & Encodes the harmful request and asks the model to decode or operate on the encoded content. \\
Roleplay~\cite{shen2024anything} & Frames the request through a persona, fictional role, or simulated scenario to weaken refusal behavior. \\
Prefix Injection~\cite{wei2023jailbroken} & Prepends instructions that attempt to override the model's safety policy or system behavior. \\
Refusal Suppression~\cite{wei2023jailbroken} & Explicitly asks the model not to refuse, apologize, moralize, or mention safety constraints. \\
JSON Format~\cite{yoosuf2025structtransform} & Requests the answer in a structured format, such as JSON fields, to reduce natural-language refusal cues. \\
Affirmative Prefix~\cite{zhu2024advprefix} & Starts the assistant response with an affirmative phrase so generation continues in a compliant direction. \\
Past Tense~\cite{andriushchenko2025does} & Rewrites the harmful instruction as a past-tense or historical description. \\
Low-resource Language~\cite{huang2026obscure} & Translates or obfuscates the request through a lower-resource language or multilingual form. \\
Style Transfer~\cite{luo2026simple} & Asks the model to express the harmful content in a specific style, genre, or transformed surface form. \\
Optimized Suffix~~\cite{zou2023universal,zhou2025don,zhu2024autodan} & Appends an optimized suffix or model-specific continuation pattern to induce harmful compliance. \\
\bottomrule
\end{tabularx}
\end{table*}

\section{Baseline and Solver Details}
\label{app:baseline_details}

\paragraph{Attribution baselines.}
ASR ranks attacks by no-defense attack success rate and does not query the defender-side utility function.
AOO measures empty-context marginal gain,
\[
\mathrm{AOO}_i=v_\theta(\{a_i\})-v_\theta(\emptyset),
\]
while LOO measures full-context marginal loss,
\[
\mathrm{LOO}_i=v_\theta(\mathcal{A})-v_\theta(\mathcal{A}\setminus\{a_i\}).
\]
AttackSHAP averages marginal utility over all subset contexts.
In synthetic experiments, reference AttackSHAP is computed exactly from the full utility table.
In surrogate-based estimation, AttackSHAP is computed by replacing true utility queries with \(\hat v_\theta\).

\paragraph{Subset-selection baselines.}
For budget \(k\), \textsc{Random} averages over random size-\(k\) subsets.
\textsc{TopK-ASR}, \textsc{TopK-AOO}, \textsc{TopK-LOO}, and \textsc{TopK-AttackSHAP} rank individual attacks by the corresponding score and return the top \(k\).
\textsc{A-MESS-Greedy} adaptively queries true utilities and adds the attack with the largest observed marginal gain at each step.
\textsc{A-MESS-Surrogate} first learns \(\hat v_\theta\) from sampled subset utilities and then searches the learned utility landscape.
When the full synthetic landscape is available, \textsc{OracleSearch} exhaustively searches all size-\(k\) subsets and is used only as an upper bound.

\paragraph{Synthetic surrogate training.}
In the synthetic experiments, A-MESS-Surrogate is trained under utility-query budgets \(M\in\{100,300,500,700,900\}\).
For each task and budget, models are selected by validation error, and the selected surrogate is used for downstream AttackSHAP estimation and subset search.
Because the synthetic utility table is fully enumerated, held-out evaluation can be performed on subsets that were never used for surrogate training.

\end{document}